\documentclass[journal]{IEEEtran}

\usepackage{amssymb}
\usepackage{graphicx}
\usepackage{color}
\usepackage{soul} 
\usepackage{mathtools}
\usepackage{url}
\usepackage{caption}
\usepackage{subcaption}
\usepackage{multirow}
\usepackage{xcolor}
\usepackage[colorlinks = true,
            linkcolor = blue,
            urlcolor  = blue,
            citecolor = blue,
            linkbordercolor = blue,
            anchorcolor = blue]{hyperref}

\graphicspath{{Figures/}}

\ifCLASSINFOpdf
\else
\fi

\begin{document}

\title{Adaptive Multi-Class Audio Classification in Noisy In-Vehicle Environment}

\author{Myounggyu~Won$^{1}$, Haitham~Alsaadan$^{1}$, and Yongsoon Eun$^2$\\
$^1$WENS Lab, South Dakota State University, Brookings, SD, United States\\
$^2$CPS Global Center, Daegu Gyeongbuk Institute of Science and Technology, Daegu, South Korea\\
\{myounggyu.won,haitham.alsaadan\}@sdstate.edu, yeun@dgist.ac.kr}%



\maketitle

\begin{abstract}
With ever-increasing number of car-mounted electric devices and their complexity, audio classification is increasingly important for the automotive industry as a fundamental tool for human-device interactions. Existing approaches for audio classification, however, fall short as the unique and dynamic audio characteristics of in-vehicle environments are not appropriately taken into account. In this paper, we develop an audio classification system that classifies an audio stream into music, speech, speech+music, and noise, adaptably depending on different driving environments. A case study is performed with four different driving environments, \emph{i.e.,} highway, local road, crowded city, and stopped vehicle. More than 420 minutes of audio data including various genres of music, speech, speech+music, and noise are collected from the driving environments. The results demonstrate that the proposed approach improves the average classification accuracy up to 166\%, and 64\% for speech, and speech+music, respectively, compared with a non-adaptive approach in our experimental settings.
\end{abstract}


\IEEEpeerreviewmaketitle

\section{Introduction}
\label{sec:introduction}

Audio classification has been primarily used for automated multimedia content analysis~\cite{xie2011pitch} and is rapidly expanding its range of applications into diverse areas. It is a key front-end technology for speech recognition algorithms that are fundamental technological ingredients to help advance the smart city initiative by facilitating human-computer interactions. For example, it is increasingly applied to in-vehicle intelligent transportation systems for safer and more efficient control of ever increasing car-mounted electronic devices and integrated smartphones. An example is the automated vehicle audio volume control system that automatically adjusts the volume depending on on-going human conversation.

Numerous solutions have been proposed for audio classification, especially concentrating on classifying audio data into music and speech. Lu~\emph{et al.} presented a k-nearest neighbor (KNN)-based classification algorithm along with some new features like the noise frame ratio and band periodicity~\cite{lu2001robust}. Chen~\emph{et al.} utilized support vector machine (SVM) to extract audio data from movies~\cite{chen2006mixed}. They selected four major features, \emph{i.e.,} silence ratio, and variance of zero-crossing rate from the time domain, and sub-band energy, and harmonic ratio from the frequency domain. Dogan~\emph{et al.} adopted a decision-tree for classification based on MPEG-7 features including audio spectrum centroid, and audio power~\cite{dogan2009content}. Xie \emph{et al.} introduced two new pitch-density-based features namely the average pitch-density and relative tonal power density for more effective classification~\cite{xie2011pitch}\cite{fu2009noise}. However, these approaches do not take into account the unique and dynamic audio characteristics of in-vehicle environments, and their effect on the classification accuracy.

The key idea of the proposed system is that the classification accuracy is improved by individually training classification models depending on driving environments, and by developing a new classification algorithm that effectively and adaptively utilizes those heterogeneous classification models. Consequently, in this paper, we propose to design, implement, and evaluate a multi-class audio classification system that classifies an input audio stream into four different audio types, \emph{i.e.,} music, speech, speech+music, and noise, by selecting an appropriate classification model depending on diverse driving environments.

The proposed system consists of a feature selection module that identifies an optimal feature set, and a classification algorithm based on the support vector machine (SVM) that adapts to varying driving environments. The training phase of the proposed system allows users to build classification models for a wide range of driving environments, \emph{e.g.,} different weather conditions, vehicle models, and road conditions. In this paper, we perform a case study with four different driving environments, \emph{i.e.,} highway, local road, city, and stopped vehicle. The results indicate that accuracy improvements of up to 166\%, and 64\% are achieved for identification of speech, and speech+music, respectively, compared with a non-adaptive solution (the accuracy for music was high in both adaptive and non-adaptive approaches). The results also show that the proposed system improves the accuracy compared with an existing classification algorithm~\cite{xie2011pitch}.

The contributions of this paper are summarized as follows.

\begin{itemize}
  \item We provide a corpus of more than 420 mins of real-world audio data (\emph{i.e.,} speech, music, speech+music, and noise) collected from various driving environments (\emph{i.e.,} highway, local road, city, and stopped vehicle).
  \item We identify that non-adaptive classification approaches face significant performance degradation if we do not account for varying driving environments.
  \item We develop an adaptive multi-class audio classification system that adaptively applies classification models depending on driving environments to achieve high classification accuracy.
  \item We perform extensive experiments to validate the effectiveness of the proposed approach.
\end{itemize}

This paper is organized as follows. In Section~\ref{sec:motivation}, we analyze the audio characteristics of different driving environments and identify the performance degradation of existing approaches. We then describe the design of the proposed system in Section~\ref{sec:system_design}. Experimental results are presented in Section~\ref{sec:experiment}, and we conclude in Section~\ref{sec:conclusion}.

\section{Motivation}
\label{sec:motivation}

This section analyzes the classification accuracy of traditional non-adaptive approaches to justify the need for an adaptive solution for in-vehicle environments. More specifically, supervised learning was adopted to create classification models for speech, music, speech+music, and noise using audio data collected from a stopped vehicle, whereas the test data were obtained from varying driving environments, \emph{i.e.,} highway, local road, and city. The classification models were then used to classify the collected test data.

\begin{figure}[!htbp]
\centering
\includegraphics[width=0.6\columnwidth]{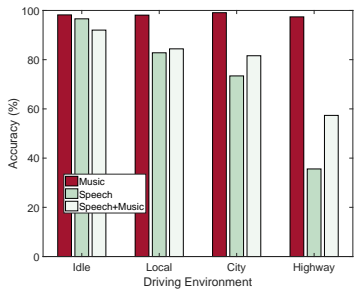}
\caption {The average accuracy of a non-adaptive approach for varying driving environments.}
\label{fig:motivation}
\end{figure}

Figure~\ref{fig:motivation} depicts the average classification accuracy for different audio types collected from varying driving environments. The accuracy for music was fairly high in all test environments. However, the accuracy for speech and speech+music was significantly impacted by different driving environments. 

\begin{figure}[!htbp]
\centering
\includegraphics[width=0.99\columnwidth]{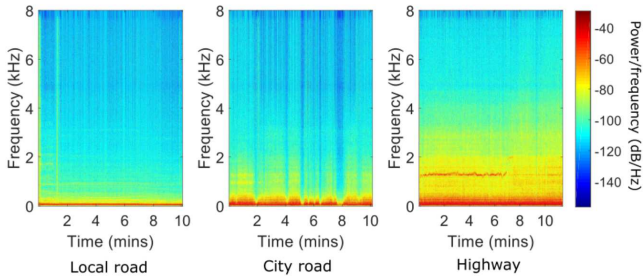}
\caption {Spectrogram of in-vehicle noise in different driving environments.}
\label{fig:spectrogram}
\end{figure}

This significant accuracy degradation is primarily attributed to substantially different noise characteristics of different driving environments. The spectrogram of in-vehicle noise for varying driving environments shows that the frequency range as well as the intensity of noise were unique in each driving environment (Figure~\ref{fig:spectrogram}). In addition, the wide frequency range of noise that spans up to 4Khz implies that applying a simple bandpass filter is not enough to completely remove the noise in that typical human voice has a frequency range between 300Hz and 3Khz~\cite{power2003psychology}.

Motivated by these experimental observations, we propose that, to minimize the performance degradation, classification models need to be generated separately for varying driving environments, and adaptively applied depending on the current driving environment. More specifically, we consider four different driving environments, \emph{i.e.,} local road, city, highway, and stopped vehicle. However, it is worth to mention that it can be easily extended to incorporate other definitions of driving environments, \emph{e.g.,} weather conditions, vehicle models, and road conditions. In the following sections, we present the details of the proposed adaptive audio classification system.

\section{System Design}
\label{sec:system_design}

This section presents an overview of the proposed system, followed by the details of the major components of the system.

\subsection{Overview}

\begin{figure}[!htbp]
\centering
\includegraphics[width=0.99\columnwidth]{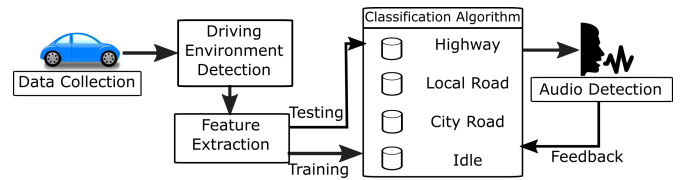}
\caption {System overview.}
\label{fig:system_overview}
\end{figure}

Figure~\ref{fig:system_overview} illustrates an overview of the proposed system. The driving environment detection module identifies the current driving environment. Driving environments can be identified using various methods, \emph{e.g.}, based on vehicle locations utilizing the GPS sensor, or collected audio data may be used since driving environments have distinctive audio characteristics. Collected audio data then go through the feature extraction module where appropriate features are extracted from the data. There are two system modes: training and testing. In the training mode, extracted features are used to train classification models corresponding to the current driving environment, whereas, in the testing mode, extracted features are provided to the classification algorithm that classifies the input data into music, speech, speech+music, or noise according to the current driving environment. Once classification is completed, input data are saved so that they are used to consolidate the classification models.

\subsection{Feature Extraction}
\label{sec:feature_extraction}

\begin{table}
\caption{Selected features.}
\begin{center}
\begin{tabular}{l*{5}{l}}
Type              & Feature \\
\hline\hline
Time domain & Root mean square \\
 & Zero-crossing rate \\
 & High zero-crossing rate ratio \\
 & Low short time energy ratio \\
 & Noise frame ratio \\
 & Silence frame ratio \\ \hline
Spectral domain & Spectral centroid \\
 & Spectral spread \\
 & Spectral flux \\
 & Spectral kurtosis \\
 & Spectral roll-off frequency \\
 & Band period \\
 & Subband energy distribution \\
 & Mel frequency cepstral coefficients \\
 & Linear predictive cepstral coefficients \\
 & Linear spectral pairs \\ \hline\hline
\end{tabular}
\end{center}
\end{table}

It is of paramount importance to select appropriate features to achieve high classification accuracy~\cite{xie2011pitch}. We consider 16 primary features including well-known and effective features for audio classification into speech, music, and and environmental sound~\cite{lu2001robust}. These features that are extracted using MirToolbox~\cite{lartillot2008matlab} are summarized in Table I. The wrapper-based feature selection method is then used to find a feature set that optimizes the classification accuracy for each classifier that we use in our classification algorithm~\cite{guyon2003introduction}. The details of the classification algorithm is described in the following section.

\subsection{Classification Algorithm}
\label{sec:classification}

Extracted features are used to train classification models for each audio type and for each driving environment. Support vector machine (SVM) is known to show superior performance compared with other classifiers~\cite{lu2002content}\cite{xie2011pitch}. We adopt SVM to create the models based on supervised learning.

\begin{figure}[!htbp]
\centering
\includegraphics[width=0.6\columnwidth]{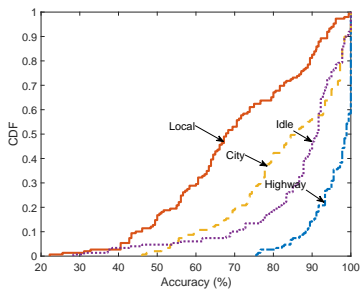}
\caption {CDF of accuracy for music.}
\label{fig:sm_cdf_test}
\end{figure}

A challenge is that audio data need to be classified into multiple audio types using SVM that is inherently a binary classifier. A known approach to implement multi-class audio classification is to first classify input data into speech and non-speech~\cite{xie2011pitch}. To do this, audio types of speech and speech+music are combined to create the `speech' class, and the combination of music, and environmental sound is used to build the `non-speech' class. Once input data is determined to be speech or non-speech, the input data is subsequently classified into speech, speech+music, music, and environmental sound using binary classifiers. However, when this approach was applied to our data set, the accuracy for music was significantly degraded as shown in the cumulative distribution function (CDF) graph of the music accuracy for different driving environments (Figure~\ref{fig:sm_cdf_test}), potentially due to the `combination effect' of multiple audio data types.

\begin{figure}[!htbp]
\centering
\includegraphics[width=0.7\columnwidth]{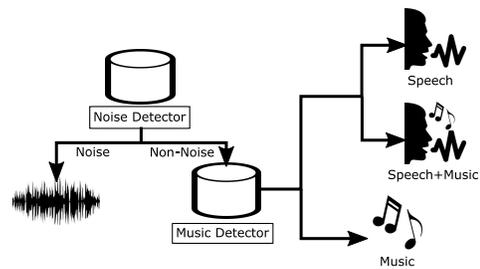}
\caption {Overview of classification algorithm.}
\label{fig:overview}
\end{figure}

In order to address this issue of degraded accuracy for music, a new classification algorithm is proposed. The proposed algorithm is displayed in Figure~\ref{fig:overview}. It consists of two modules: Noise Detector, and Music Detector. The noise detector performs testing of input data using three types of classifiers, \emph{i.e.,} `speech vs noise', `music vs noise', and `speech+music vs noise'. Essentially they are a set of one-versus-one classifiers that are specifically focused on identifying noise. Each classifier determines that the input data is noise if the accuracy is greater than a threshold. Consequently, the noise detector specifies that the input data is noise if all three classifiers indicate that it is noise. Once the noise detector finds that input data is non-noise, \emph{i.e.,} speech, speech+music, or music, the music detector starts to run. The design of music detector is similar to the noise detector; the difference is that it has one-versus-one classifiers that are used to discern music. More specifically it tests the input data stream against two types of classifiers `speech vs music', and `speech+music vs music' to identify music. Finally, if the result is non-music, a binary classifier is used to differentiate speech, or speech+music.

\section{Experimental Results}
\label{sec:experiment}

In this section, we first measure the running time of the proposed system to understand its potential for being integrated with a real-time application, \emph{e.g.,} a mobile App. Next, the accuracy of the proposed system is measured. Lastly, we evaluate the effect of varying music genres on the classification accuracy.

\subsection{Experimental Setup}
\label{sec:experimental_setup}

Audio data are collected from cities of Brookings (Local), Sioux Falls (City), Interstate-29 (Highway), and a stopped vehicle in South Dakota, United States. More specifically, 10mins of audio data for each type were collected, \emph{i.e.,} speech, music (4 different genres), speech+music, and noise from varying driving environments, totalling 280mins of data. For testing, 5mins of audio data were collected for each audio type and for each driving environment. The total data size was about 420mins. The audio streams were 16kHz mono with 16bit per sample. Audio streams were divided into non-overlapping frames of 100ms, and for analysis with longer frames, clips of 1 second were used.

\subsection{Execution Time}
\label{sec:running_time}

\begin{figure}[!htbp]
\begin{minipage}[b]{0.48\columnwidth}
\centering
\includegraphics[width=\columnwidth]{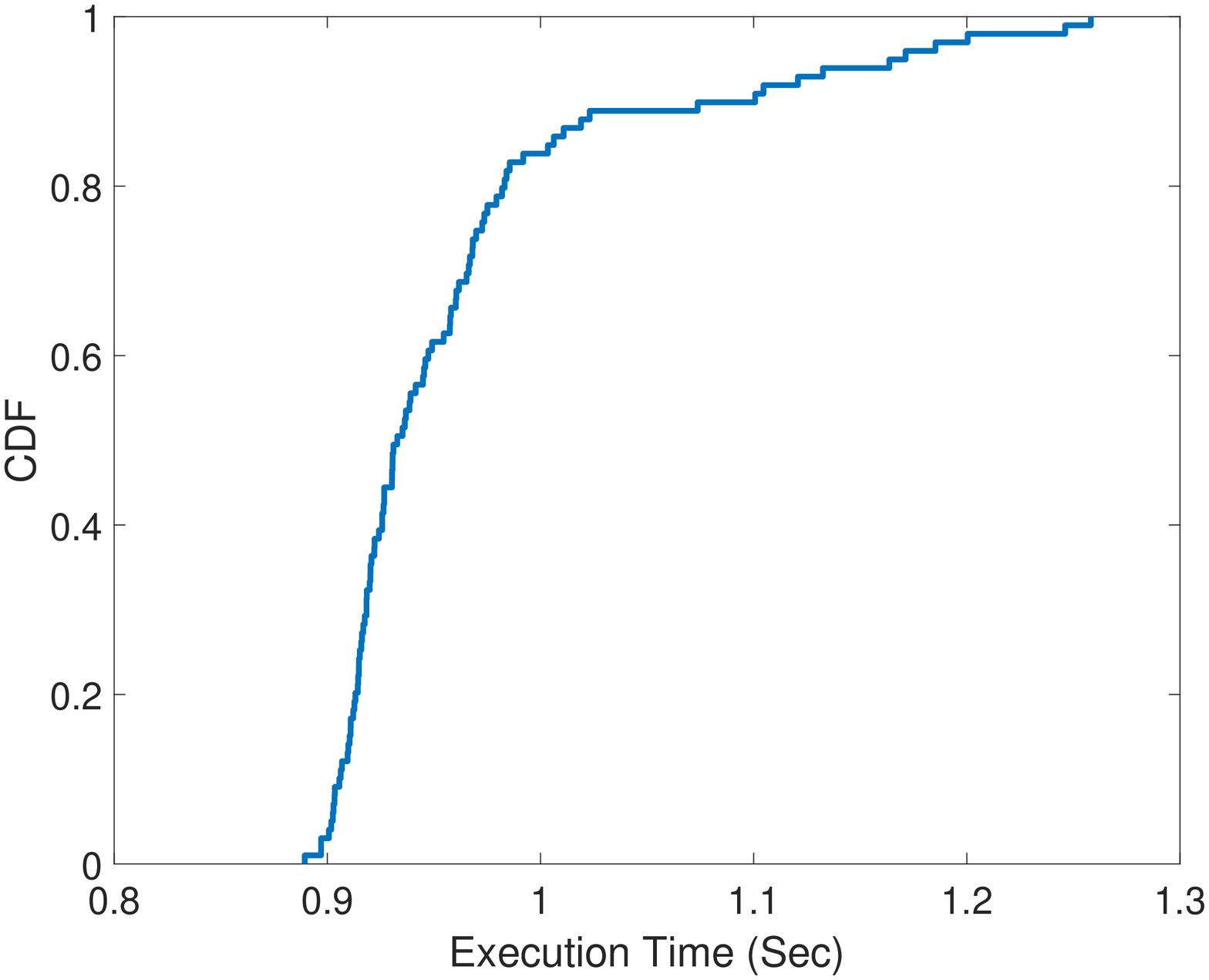}
\caption {CDF of execution time.}
\label{fig:feature_speed_cdf}
\end{minipage}
\hspace{1mm}
\begin{minipage}[b]{0.48\columnwidth}
\centering
\includegraphics[width=\columnwidth]{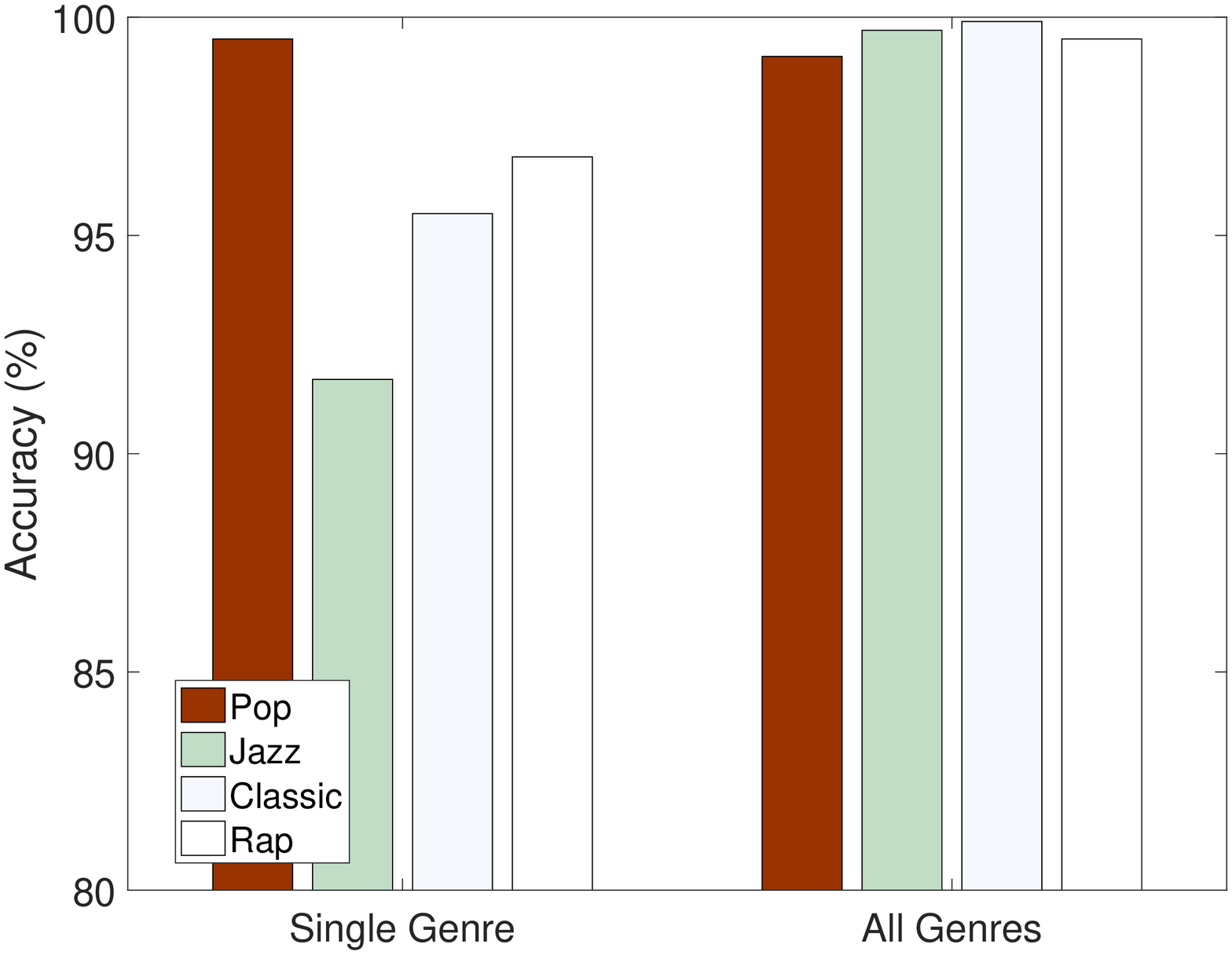}
\caption {Accuracy for music with different genres.}
\label{fig:adaptive_results}
\end{minipage}
\end{figure}


The system execution time is important in applying the proposed system to real-time applications such as mobile apps for smart transportation. We measured the execution time for 1-second clips. The results are depicted in Figure~\ref{fig:feature_speed_cdf}. The CDF graph indicates that, in most cases, the classification was completed in around 1 second which makes feasible to perform classification approximately every 2 seconds.

\subsection{Classification Accuracy}
\label{sec:accuracy}

\begin{table}[!htbp]
\caption{Accuracy of the proposed system.}
\begin{center}
\begin{tabular}{l*{5}{c}}
Category              & Highway & Local & City  & Idle \\
\hline\hline
Speech & 92.3\% & 91.4\% & 96.5\% & 96.3\% \\
Music &  97.1\% & 92.9\% & 97.8\% &  98.2\% \\
Speech+Music & 93.5\% & 89.8\% & 88.7\% & 92.0\% \\
Noise & 98.4\% & 99.1\% & 99.8\% & 99.7\% \\ \hline
Average & 95.3\% & 93.3\% & 95.7\% & 96.6\% \\  \hline\hline
\end{tabular}
\end{center}
\end{table}

This section analyzes the accuracy of the proposed system, and compare the results with a non-adaptive approach. The results are also compared with a different classification algorithm~\cite{xie2011pitch}. The average accuracy of the proposed adaptive approach is shown in Table II. As shown, significant improvement in accuracy was achieved for all driving environments. The highest accuracy was achieved for the idle environment (stopped vehicle), but the highest improvement in accuracy was achieved for the highway. More specifically, the accuracy for highway was improved by 166\% and 64\% for speech, speech+music, respectively, compared with a non-adaptive approach, whereas the accuracy was not improved much for music; in local road, the accuracy was even slightly decreased. An interesting observation was that no correlation between the noise level of driving environments and the accuracy was found after applying the adaptive method. It was also observed that the noise was successfully detected with the average accuracy over 93\% in all driving environments. Overall, with the proposed adaptive approach, the accuracy was high in all driving environments.

\begin{table}[!htbp]
\caption{Accuracy of~\cite{xie2011pitch} for in-vehicle environment. }
\begin{center}
\begin{tabular}{l*{5}{c}}
Category           & Highway & Local & City  & Idle \\ \hline\hline
Speech            & 94.9\% & 94.5\% & 92.8\% & 96.3\%\\
Music               & 96.0\% & 69.1\% & 82.9\% &  86.2\% \\
Speech+Music & 89.7\% & 89.0\% & 85.7\% & 94.7\% \\
Noise               & 98.4\% & 99.1\% & 99.8\% & 99.7\% \\ \hline
Average           & 94.8\% & 87.9\% & 90.3\% & 94.2\% \\  \hline\hline
\end{tabular}
\end{center}
\end{table}

We also obtained the average accuracy for other classification algorithm~\cite{xie2011pitch}, which is shown in Table III -- note that the noise was detected using the same method. In comparison with the results of the proposed classification algorithm, it was observed that the accuracy for music was significantly low, which affected the average accuracy. Overall the proposed system achieved 0.5\%, 6.1\%, 6\%, and 2.6\% higher accuracy compared with Xie \emph{et al.}~\cite{xie2011pitch} in our experimental settings.




\subsection{Effect of Music Genre}
\label{sec:classifier}


We originally performed supervised training with only a single music genre, \emph{i.e.,} pop music. In order to further increase the performance, we trained the classification model for music taking into account different music genres (\emph{i.e.,} jazz, classic, and rap) and measured the accuracy. For this set of experiments, we used the highway data. The results are depicted in Figure~\ref{fig:adaptive_results}. As shown, when it is trained with a single musical genre, the accuracy was degraded for other music genres. It was observed that after training with multiple music genres, the accuracy was increased by 8\%, 5\%, and 3\% for jazz, classic, and rap, respectively.

\section{Conclusion}
\label{sec:conclusion}

We presented an adaptive multi-class music classification system for dynamic in-vehicle environments. A novel classification algorithm based on effective feature selection allows for high speech detection accuracy for driving environments with varying noise characteristics. Future work is to incorporate other useful features such as pitch-density-based features~\cite{xie2011pitch}, and bottleneck features extracted using deep neural network~\cite{zhang2016deep}.

\section*{Acknowledgment}
This research was supported in part by Global Research Laboratory Program (2013K1A1A2A02078326) through NRF, and DGIST Research and Development Program (CPS Global Center) funded by the Ministry of Science, ICT \& Future Planning.

\ifCLASSOPTIONcaptionsoff
  \newpage
\fi

\bibliographystyle{IEEEtran}
\bibliography{reference_audio}

\end{document}